\documentclass[12pt,showpacs,showkeys,amsmath,amssymb]{revtex4}
\usepackage{amsmath,amsfonts,amsthm,amscd,amssymb,latexsym}
\usepackage{bm}
\usepackage{dcolumn}
\usepackage{graphicx}
\usepackage{epstopdf}
\usepackage{color}
\usepackage{epsf}
\usepackage{epsfig}
\usepackage{graphicx, epic, eepic, color}

\date{\today}

\begin{document}
\title{Nonlocal General Relativity\footnote{Dedicated to the memory of M. Hossein Partovi (1941-2014).}}

\author{B. Mashhoon}
\email{mashhoonb@missouri.edu}
\affiliation{Department of Physics and Astronomy,
University of Missouri, Columbia, Missouri 65211, USA}

\begin{abstract} 
 A brief account of the present status of the recent nonlocal generalization of Einstein's theory of gravitation is presented. The main physical assumptions that underlie this theory are described. We clarify the physical meaning and significance of Weitzenb\"ock's torsion, and emphasize its intimate relationship with the gravitational field, characterized by the Riemannian curvature of spacetime. In this theory, nonlocality can simulate dark matter; in fact, in the Newtonian regime, we recover the phenomenological Tohline-Kuhn approach to modified gravity. To account for the observational data regarding dark matter, nonlocality is associated with a characteristic length scale of order 1 kpc. The confrontation of nonlocal gravity with observation is  briefly discussed.
\end{abstract}



\maketitle

\section{Introduction}

In relativity theory, Lorentz transformations are first extended in a \emph{pointwise} manner to accelerated systems in special relativity and then to gravitational fields in general relativity via Einstein's \emph{local} principle of equivalence~\cite{Ei, Mas}. In this approach, the first step is based on the assumption that an accelerated observer in Minkowski spacetime, at each event along its world line,  is physically equivalent to an otherwise identical momentarily comoving inertial observer. This \emph{hypothesis of locality} generally amounts to a first approximation, since field measurements cannot be performed instantaneously and require an extended time interval. To go beyond the locality postulate, one must take the past history of the accelerated observer into account. Such history-dependent theories are nonlocal~\cite{BM}. Nonlocal theories of special and general relativity have recently been developed~\cite{M1, NL1, NL2, NL3, NL4, NL5, NL6, NL7, NL8, NL9}. It turns out that nonlocal general relativity simulates dark matter; that is, according to this theory what appears as dark matter in astrophysics is essentially a manifestation of the nonlocality of the gravitational interaction. 

Lorentz invariance is a fundamental symmetry of nature and involves the relationship between the measurements of ideal inertial observers in uniform relative motion in Minkowski spacetime. Such hypothetical observers do not truly exist; indeed, actual observers are all more or less accelerated. Thus a physical hypothesis is required to relate the measurements of actual accelerated observers with those of hypothetical momentarily comoving inertial observers. The special theory of relativity is based on the hypothesis of locality, which postulates that an accelerated observer is pointwise inertial, so that Lorentz transformations may be applied event by event along its world line to determine what the accelerated observer measures. This locality postulate fits in well with Einstein's local principle of equivalence, since these together imply that in general relativity (GR) an observer in a gravitational field is pointwise inertial~\cite{Mas}.

The hypothesis of locality has its roots in the Newtonian mechanics of classical point particles. Thinking of classical physics in terms of particles and waves, it is clear that the locality postulate is valid for point particles and leads to a theory of \emph{pointlike coincidences}, whereas the measurement of wave properties requires an extended period of time. Indeed, Bohr and Rosenfeld~\cite{BR} have shown that the measurement of the classical electromagnetic field cannot be done instantaneously and necessitates a certain spacetime averaging procedure over past events. Thus to go beyond the hypothesis of locality for the measurement of classical fields, the past history of the accelerated observer and the fields must be taken into account. To incorporate history dependence, the usual partial differential equations for the fields must be replaced by integro-differential equations. In this way, one is led to nonlocal special relativity, in which fields are local, but satisfy integro-differential field equations~\cite{M1}.

Is gravitation history dependent? Einstein interpreted the principle of equivalence of inertial and gravitational masses to mean that an intimate connection exists between inertia and gravitation. We follow Einstein's interpretation, but do not postulate a local equivalence between inertia and gravitation as in GR; rather, we wish to extend GR to make it history dependent along the same lines as in nonlocal special relativity. That is, GR should become history dependent, since accelerated systems in Minkowski spacetime are expected to be history dependent. It turns out that within the framework of teleparallelism~\cite{BH, AP, Ma}, GR has an equivalent tetrad formulation, namely, GR$_{||}$, that can be rendered nonlocal in close analogy with the nonlocal electrodynamics of media~\cite{NL1, NL2}. To implement these ideas, we need an extended general relativistic framework in which the Riemannian metric is supplemented with \emph{two} metric-compatible connections, namely, the standard Levi-Civita connection (${^0}\Gamma^\mu_{\alpha \beta}$) and the Weitzenb\"ock connection ($\Gamma^\mu_{\alpha \beta}$). A left superscript ``0" is used throughout to denote geometric quantities in GR that are directly related to the Levi-Civita connection.  
We describe this extended tetrad framework in section II. Section III is devoted to the formulation of nonlocal gravity (NLG). Observational aspects of NLG are treated in section IV. Section V contains a discussion of our results. 

It is important to note that there are other approaches to nonlocal gravitation---see Refs.~\cite{TW, BMMS} and the references cited therein; however, what sets our specific approach completely apart from other nonlocal modifications of GR is the initial physical motivation stemming from the hypothesis of locality and the specific path followed from nonlocal special relativity to NLG.

\section{EXTENSION of GR FRAMEWORK}

In general relativity, spacetime is a smooth four-dimensional manifold with a Lorentz metric such that the invariant spacetime interval $ds$ is given by
\begin{equation}\label{J1}
ds^2=g_{\mu \nu}~ dx^\mu \, dx^\nu\,.
\end{equation}
The path of a test particle of constant inertial mass $m$ is obtained as usual from
\begin{equation}\label{J2}
\delta \int -m c~ds=0\,,
\end{equation}
which results in the geodesic equation of motion for the test particle
\begin{equation}\label{J3}
\frac{d^2 x^{\mu}}{ds^2}+{^0}\Gamma^\mu_{\alpha \beta}~ \frac{dx^\alpha}{ds} \frac{dx^\beta}{ds} = 0\,.
\end{equation}
Similarly, rays of radiation follow null geodesics of  the spacetime manifold.  The four-velocity vector of a test particle, $u^\mu=dx^\mu/ds$, is parallel transported along a geodesic via the Levi-Civita connection that is given by the Christoffel symbols  
\begin{equation}\label{J4}
{^0}\Gamma^\mu_{\alpha \beta}= \frac{1}{2} g^{\mu \nu} (g_{\nu \alpha,\beta}+g_{\nu \beta,\alpha}-g_{\alpha \beta,\nu})\,.
\end{equation}
This symmetric connection is torsion-free but has Riemannian curvature (${^0}R_{\mu \nu \rho \sigma}$). In our convention, an event in spacetime has coordinates $x^\mu=(ct, x^i)$, where Greek indices run from 0 to 3, while Latin indices run from 1 to 3; moreover, the spacetime metric has signature +2, $\kappa:=8 \pi G/c^4$ and $c=1$, unless otherwise specified. The gravitational field equations in GR are given by~\cite{Ei} 
\begin{equation}\label{J5}
 {^0}R_{\mu \nu}-\frac{1}{2} g_{\mu \nu}\,{^0}R+ \Lambda\, g_{\mu \nu}= \kappa~T_{\mu \nu}\,,
 \end{equation}
where ${^0}R_{\mu \nu}={^0}R^{\alpha}{}_{\mu \alpha \nu}$ is the symmetric Ricci tensor, ${^0}R= g^{\mu \nu}~{^0}R_{\mu \nu}$ is the scalar curvature, $\Lambda$ is the cosmological constant and $T_{\mu \nu}$ is the \emph{symmetric} energy-momentum tensor of matter.

Each observer in spacetime carries an orthonormal tetrad frame $\lambda^\mu{}_{\hat {\alpha}}(x)$, where $\lambda^\mu{}_{\hat {0}}$ is the observer's unit temporal direction and $\lambda^\mu{}_{\hat {i}}$\,, $i=1, 2, 3,$ constitute its local spatial frame. The projection of tensor fields on an observer's tetrad frame indicates the local measurement of the corresponding physical quantities by the observer.  Spacetime indices are raised and lowered via the metric tensor $g_{\mu \nu}$, while the hatted tetrad indices---that is, the local Lorentz indices---are raised and lowered via the Minkowski metric tensor $\eta_{\mu \nu}$ given by diag$(-1,1,1,1)$ in our convention.  The orthonormality condition for the tetrad frame $\lambda^\mu{}_{\hat {\alpha}}(x)$ can be expressed as 
\begin{equation}\label{J6}
 g_{\mu \nu}(x)= \eta_{\hat {\alpha} \hat  {\beta}}\,
\lambda_\mu{}^{\hat {\alpha}}(x)~ \lambda_\nu{}^{\hat {\beta}}(x)\,, 
\end{equation}
so that we can write Eq.~\eqref{J1} as 
\begin{equation}\label{J7}
 ds^2= \eta_{\hat {\alpha} \hat  {\beta}}\,~dx^{\hat {\alpha}}\, dx^{\hat  {\beta}}\,,
\end{equation}
where $dx^{\hat {\alpha}}=\lambda_\mu{}^{\hat {\alpha}}~dx^\mu$. Thus the tetrad provides the local connection between spacetime quantities and local Lorentz quantities for the observer. 

The differential form $dx^{\hat {\alpha}}=\lambda_\mu{}^{\hat {\alpha}}~dx^\mu$ is in general not exact, since an exact form would imply that it is integrable, namely, that there exist four functions $x^{\hat {\alpha}}(x)$ such that $\lambda_\mu{}^{\hat {\alpha}}=\partial x^{\hat {\alpha}}/\partial x^\mu$. It would then follow from Eq.~\eqref{J6} that we are in Minkowski spacetime with ${^0}R_{\mu \nu \rho \sigma}=0$. Indeed, the family of observers with such a frame field would be the \emph{static} inertial observers with tetrad frames that are all  parallel and point along the Cartesian coordinate axes of a global inertial frame with coordinates $x^{\hat {\alpha}}$. If ${^0}R_{\mu \nu \rho \sigma} \ne 0$, then $dx^{\hat {\alpha}}$ is not integrable and, at each event, such 1-forms will constitute a noncoordinate or anholonomic Lorentz basis. Therefore, to change a holonomic spacetime index of a tensor into an anholonomic local Lorentz index or vice versa, one can simply project the tensor onto the corresponding local tetrad frame.

In GR, the gravitational field is identified with spacetime curvature; moreover, one traditionally works with admissible coordinate systems~\cite{Sy, BCM}. Coordinate bases are \emph{holonomic}, while noncoordinate bases are \emph{anholonomic}. In differential geometry, one can work with either holonomic or anholonomic bases. We find it convenient to work primarily with holonomic bases in this paper. 

In a patchwork of admissible coordinate charts in spacetime, consider a smooth orthonormal tetrad frame field $e^\mu{}_{\hat{\alpha}}(x)$ corresponding to a \emph{preferred} set of observers. We use this preferred tetrad system to define a new linear \emph{Weitzenb\"ock connection}~\cite{We}
\begin{equation}\label{J8}
\Gamma^\mu_{\alpha \beta}=e^\mu{}_{\hat{\rho}}~\partial_\alpha\,e_\beta{}^{\hat{\rho}}\,.
\end{equation}
It can be checked directly that this nonsymmetric connection is curvature-free; moreover, it is so constructed as to render the frame field parallel, namely, $\nabla_\nu\,e_\mu{}^{\hat{\alpha}}=0$, where $\nabla_\nu$ here denotes covariant differentiation with respect to the Weitzenb\"ock connection. This circumstance leads to \emph{teleparallelism}; that is, distant vectors can be considered parallel if they have the same components with respect to their local preferred frames. The Levi-Civita and 
Weitzenb\"ock connections are both compatible with the spacetime metric tensor; indeed, the latter is a consequence of $\nabla_\nu\, g_{\alpha \beta}=0$, which follows from the orthonormality relation $g_{\mu \nu} = e_\mu{}^{\hat{\alpha}}~ e_\nu{}^{\hat{\beta}}~ \eta_{\hat{\alpha} \hat{\beta}}$. 

Under a general transformation of coordinates $x^\mu \mapsto x'^\mu$, a linear connection transforms  just as, say,  the Weitzenb\"ock connection 
\begin{equation}\label{J9}
\Gamma'^\mu_{\alpha \beta}=\frac{\partial x'^\mu}{\partial x^\nu}\frac{\partial x^\gamma}{\partial x'^\alpha}\frac{\partial x^\delta}{\partial x'^\beta}~\Gamma^\nu_{\gamma \delta}+\frac{\partial x'^\mu}{\partial x^\nu}~\frac{\partial^2 x^\nu}{\partial x'^\alpha \partial x'^\beta}\,.
\end{equation}
Therefore, the difference between two linear connections on the same spacetime manifold is a tensor. In this way, we have the \emph{torsion} tensor 
\begin{equation}\label{J10}
 C_{\alpha \beta}{}^{\mu}=\Gamma^{\mu}_{\alpha \beta}-\Gamma^{\mu}_{\beta \alpha}=e^\mu{}_{\hat{\rho}}\Big(\partial_{\alpha}e_{\beta}{}^{\hat{\rho}}-\partial_{\beta}e_{\alpha}{}^{\hat{\rho}}\Big)\,
\end{equation}
and the \emph{contorsion} tensor
\begin{equation}\label{J11}
K_{\alpha \beta}{}^\mu= {^0} \Gamma^\mu_{\alpha \beta} - \Gamma^\mu_{\alpha \beta}\,.
\end{equation}
From the compatibility of the Weitzenb\"ock connection with the metric, namely, $\nabla_\gamma\,g_{\alpha \beta}=0$, we find 
\begin{equation}\label{J12}
 g_{\alpha \beta , \gamma}= \Gamma^\mu_{\gamma \alpha}\, g_{\mu \beta} + \Gamma^\mu_{\gamma \beta}\, g_{\mu \alpha}\,,
\end{equation}
which can be substituted in the Christoffel symbols to show that the contorsion tensor is linearly related to the torsion tensor via
\begin{equation}\label{J13}
K_{\alpha \beta \gamma} = \frac{1}{2} (C_{\alpha \gamma \beta}+C_{\beta \gamma \alpha}-C_{\alpha \beta \gamma})\,.
\end{equation}
The torsion tensor is antisymmetric in its first two indices, while the contorsion tensor is antisymmetric in its last two indices. 

There is a subtle correlation between the curvature of the Levi-Civita connection and the torsion of the 
Weitzenb\"ock connection.  To illustrate this point, let us first imagine that $C_{\alpha  \beta \gamma}=0$. This is mathematically equivalent, via Eq.~\eqref{J10}, to the requirement that $d\,(e_\mu{}^{\hat {\alpha}}~dx^\mu)=0$. On a smoothly contractible spacetime domain, every closed form is exact in accordance with the Poincar\'e lemma. In this case, there are thus four functions $x^{\hat {\alpha}}(x)$ such that $e_\mu{}^{\hat {\alpha}}~dx^\mu=dx^{\hat {\alpha}}$ or $e_\mu{}^{\hat {\alpha}}=\partial x^{\hat {\alpha}}/\partial x^\mu$. As before, it follows from the orthonormality condition that we are  back in Minkowski spacetime where our preferred observers are the static inertial observers of a global inertial frame with coordinates $x^{\hat {\alpha}}$ such that the tetrad axes are all parallel with the corresponding Cartesian coordinate axes. Therefore, $C_{\alpha  \beta \gamma}=0$ implies that ${^0}R_{\mu \nu \rho \sigma}=0$, so that there is no gravitational field. In the presence of gravitation, however, ${^0}R_{\mu \nu \rho \sigma}\ne 0$ and this implies that $C_{\alpha  \beta \gamma}\ne 0$. It thus appears that in \emph{curved} spacetime, one can characterize the gravitational field via the torsion tensor as well. 

In extended GR, the parallel frame field defined by the Weitzenb\"ock connection is the natural generalization of the parallel frames of static inertial observers in a global inertial frame in special relativity to the curved spacetime of general relativity. Let us recall that in the standard GR framework, 
 a parallel (or nonrotating) frame field may be defined via parallel (or Fermi-Walker) transport using the Levi-Civita connection along a timelike world line;  however, it \emph{cannot} in general be extended to a finite region, as this is obstructed by the Riemannian curvature of spacetime~\cite{M8}. The introduction of the Weitzenb\"ock connection remedies this situation. 
 
As is well known, in the curved spacetime of GR, at an event with coordinates $\bar{x}^\mu$  one can introduce locally geodesic coordinates in the neighborhood of $\bar{x}^\mu$ such that in the new coordinates the Christoffel symbols all vanish at $\bar{x}^\mu$ and geodesic world lines that pass through $\bar{x}^\mu$ are rendered locally straight. In a similar way, consider the coordinate transformation $x^\mu \mapsto x'^\mu$, 
\begin{equation}\label{J14}
x'^\mu = x^\mu-\bar{x}^\mu +\frac{1}{2} \big(\Gamma^{\mu}_{\alpha \beta}\big)_{\bar{x}}\, (x^\alpha-\bar{x}^\alpha) (x^\beta-\bar{x}^\beta) + ...\,.
\end{equation}
In the new coordinate system, however, only the \emph{symmetric} part of the Weitzenb\"ock connection
vanishes at $\bar{x}^\mu$ and $\Gamma^{\mu}_{[\alpha \beta]}=\frac{1}{2} C_{\alpha \beta}{}^\mu$ in general remains nonzero. In this case, the corresponding Weitzenb\"ock autoparallels passing through $\bar{x}^\mu$ are rendered locally straight. Thus at each event in our extended GR framework, the curvature and torsion tensors both characterize the gravitational field~\cite{LB}. In fact, the symbiotic relationship between the Riemann curvature and the Weitzenb\"ock  torsion of the spacetime manifold turns out to be crucial for the nonlocal generalization of GR.

We illustrate in Appendix A, via a specific example, how the Weitzenb\"ock  torsion tensor might be measured in a given gravitational field. In particular, we show that Weitzenb\"ock  torsion behaves like tidal acceleration and has dimensions of (length)$^{-1}$, while curvature has dimensions of (length)$^{-2}$. Moreover,  we show, among other things, that it is possible to introduce Fermi coordinates and tetrad frames in the neighborhood of an arbitrary timelike geodesic such that the Levi-Civita and Weitzenb\"ock connections both vanish along the timelike geodesic path. 

The Riemann curvature tensor can be expressed in terms of the Christoffel symbols and their derivatives; therefore, Eq.~\eqref{J11} can be used to write the Riemann curvature tensor in terms of the torsion tensor. After detailed but straightforward calculations, it is then possible to write the Einstein field equations~\eqref{J5} in Maxwellian form in terms of the torsion tensor. To this end, it proves useful to introduce an auxiliary torsion tensor
\begin{equation}\label{J15}
\mathfrak{C}_{\alpha \beta \gamma} :=K_{\gamma \alpha \beta}+C_\alpha\, g_{\gamma \beta} - C_\beta \,g_{\gamma \alpha}\,,
\end{equation}
where $C_\alpha:=- C_{\alpha}{}^{\beta}{}_{\beta}$ is the \emph{torsion vector}.  Furthermore, let ${\cal H}_{\alpha \beta \gamma}$ be an auxiliary field strength defined by
\begin{equation}\label{J16}
{\cal H}_{\alpha \beta \gamma}:= \frac{\sqrt{-g}}{\kappa}\,\mathfrak{C}_{\alpha \beta \gamma}\,,
\end{equation}
so that $\mathfrak{C}_{\alpha \beta \gamma}$ and ${\cal H}_{\alpha \beta \gamma}$ are both antisymmetric in their first two indices. Then, the Einstein tensor  $ {^0}G_{\mu \nu}$  can be written in the form
\begin{eqnarray}\label{J17}
 {^0}G_{\mu \nu}=-\kappa E_{\mu \nu} +\frac{\kappa}{\sqrt{-g}}~e_\mu{}^{\hat{\gamma}}\,g_{\nu \alpha}\, \frac{\partial}{\partial x^\beta}\,{\cal H}^{\alpha \beta}{}_{\hat{\gamma}}\,,
\end{eqnarray}      
where $E_{\mu \nu}$,
\begin{equation}\label{J18} 
\sqrt{-g}\, E_{\mu \nu}=C_{\mu \rho \sigma} {\cal H}_{\nu}{}^{\rho \sigma}
-\frac{1}{4}g_{\mu \nu}\,C_{\alpha \beta \gamma}{\cal H}^{\alpha \beta \gamma}\,,
\end{equation}
 turns out to be the trace-free energy-momentum tensor of the gravitational field in the new scheme.
Einstein's gravitational field equations now take the form
\begin{equation}\label{J19}
 \frac{\partial}{\partial x^\beta}\,{\cal H}^{\alpha \beta}{}_{\hat{\gamma}}=\sqrt{-g}\,e^\mu{}_{\hat{\gamma}}\,g^{\alpha \nu}\Big(T_{\mu \nu}+E_{\mu \nu}-\frac{\Lambda}{\kappa}g_{\mu \nu}\Big)\,. 
 \end{equation}
It follows from Eq.~\eqref{J19} and the antisymmetry of ${\cal H}^{\alpha \beta}{}_{\hat{\gamma}}$ in its first two indices that 
\begin{equation}\label{J20}
\frac{\partial}{\partial x^\mu}\,\Big[\sqrt{-g}~(T_{\hat{\alpha}}{}^\mu+ E_{\hat{\alpha}}{}^\mu-\frac{\Lambda}{\kappa}\,e^\mu{}_{\hat{\alpha}})\Big]=0\,.
 \end{equation}
This is the law of conservation of total energy-momentum tensor  that consists of contributions due to matter, the gravitational field and the cosmological constant. 

To summarize, within the context of GR, we have chosen a preferred frame field, which together with the corresponding  Weitzenb\"ock connection has generated a GR$_{||}$ framework that is the teleparallel equivalent of GR. In GR, the ten gravitational field equations can be used in principle to determine the ten components of the spacetime metric tensor. A tetrad frame field has, however, sixteen components, which are subject to ten orthonormality relations that, in effect, determine the metric in terms of the tetrad frame. This circumstance points to the six-fold degeneracy of GR$_{||}$. In fact, the extra six degrees of freedom are elements of the local Lorentz group; that is, the boosts and rotations that locally characterize one system of observers with respect to a fiducial system. This basic degeneracy of GR$_{||}$ will be removed in the nonlocal generalization of this theory in the next section. 

The tetrad formulation of GR goes back to Einstein's attempt at a classical unified field theory of gravitation and electromagnetism~\cite{Ein}. Later, in a purely gravitational context, M{\o}ller pointed out that the fundamental problem of gravitational energy in GR can be solved in the tetrad framework~\cite{Mo, PP}. A complete account of the various approaches to this subject can be found in Refs.~\cite{BH, AP, Ma} and the references cited therein. A detailed treatment of the approach to GR$_{||}$ adopted here can be found in the excellent review by Maluf~\cite{Ma}.

GR$_{||}$, the teleparallel equivalent of GR, is the gauge theory of the Abelian group of spacetime translations~\cite{BH, AP, Ma}. As such, the structure of GR$_{||}$ bears certain similarities with electrodynamics. For instance, Eq.~\eqref{J10} can be written as $C_{\alpha \beta}{}^{\hat{\rho}}=\partial_{\alpha}e_{\beta}{}^{\hat{\rho}}-\partial_{\beta}e_{\alpha}{}^{\hat{\rho}}$, so that for each tetrad index $\hat{\rho}= \hat{0}, \hat{1}, \hat{2}, \hat{3}$, $C_{\alpha \beta}{}^{\hat{\rho}}$ is analogous to the electromagnetic field tensor with vector potential $e_{\alpha}{}^{\hat{\rho}}$. The analogy with electrodynamics has provided the basis for the present approach to the nonlocal generalization of GR~\cite{NL1, NL2} that is described in the following section.

\section{Nonlocal Gravity (NLG)}

In the electrodynamics of media, the constitutive relation between $H_{\mu\nu} \mapsto (\mathbf{D}, \mathbf{H})$ and  $F_{\mu\nu} \mapsto (\mathbf{E}, \mathbf{B})$ could be nonlocal~\cite{Ja, LaLi}. Therefore, in the nonlocal electrodynamics of media, Maxwell's original equations remain unchanged, but the constitutive relation now involves the past history of the electromagnetic field. In GR$_{||}$, Einstein's field equations are analogous to Maxwell's  original equations and Eq.~\eqref{J16} plays the role of a \emph{local} constitutive relation. We wish to construct here  a nonlocal theory of gravitation in analogy with the nonlocal electrodynamics of media. To render observers nonlocal in a gravitational field in the same sense as in nonlocal special relativity, we simply change Eq.~\eqref{J16} to 
\begin{equation}\label{K1}
^{NLG}{\cal H}_{\mu \nu \rho}:= \frac{\sqrt{-g}}{\kappa}\,(\mathfrak{C}_{\mu \nu \rho}+ N_{\mu \nu \rho})\,,
\end{equation}
where $N_{\mu \nu \rho}=-N_{\nu \mu \rho}$ is a tensor involving the past history of the gravitational field. The field equations of nonlocal gravity are therefore obtained from Eqs.~\eqref{J18} and~\eqref{J19} by substituting $^{NLG}{\cal H}_{\mu \nu \rho}$ for ${\cal H}_{\mu \nu \rho}$. That is, the field equations of nonlocal gravity are given by
\begin{equation}\label{K2}
 \frac{\partial}{\partial x^\beta}\,\Big[\frac{\sqrt{-g}}{\kappa}\,(\mathfrak{C}^{\alpha \beta}{}_{\hat{\gamma}}+N^{\alpha \beta}{}_{\hat{\gamma}})\Big] =\sqrt{-g}\,e^\mu{}_{\hat{\gamma}}\,g^{\alpha \nu}\Big(T_{\mu \nu}+\mathcal{E}_{\mu \nu}-\frac{\Lambda}{\kappa}g_{\mu \nu}\Big)\,, 
\end{equation}
where $\mathcal{E}_{\mu \nu}$  now takes the place of $E_{\mu \nu}$, namely, 
\begin{equation}\label{K3} 
\kappa\, \mathcal{E}_{\mu \nu}=C_{\mu \rho \sigma}(\mathfrak{C}_{\nu}{}^{\rho \sigma}+N_{\nu}{}^{\rho \sigma})-\frac{1}{4}g_{\mu \nu}\,C_{\alpha \beta \gamma}(\mathfrak{C}^{\alpha \beta \gamma}+N^{\alpha \beta \gamma})\,.
\end{equation}
As before, the law of conservation of total energy-momentum tensor takes the form
\begin{equation}\label{K4}
\frac{\partial}{\partial x^\mu}\,\Big[\sqrt{-g}~(T_{\hat{\alpha}}{}^\mu+ \mathcal{E}_{\hat{\alpha}}{}^\mu-\frac{\Lambda}{\kappa}\,e^\mu{}_{\hat{\alpha}})\Big]=0\,.
\end{equation}

Let us now specify the form of the nonlocality tensor $N_{\mu \nu \rho}$. We assume, for the sake of simplicity, a nonlocal ansatz involving a \emph{scalar} kernel, namely, 
\begin{eqnarray}\label{K5}
N_{\mu \nu \rho} = - \int \Omega_{\mu \mu'} \Omega_{\nu \nu'} \Omega_{\rho \rho'}\, {\cal K}(x, x')\,X^{\mu' \nu' \rho'}(x') \sqrt{-g(x')}\, d^4x' \,,
\end{eqnarray}
where ${\cal K}(x, x')$ is the scalar \emph{causal} kernel of nonlocal gravity~\cite{NL1, NL2, NL3, NL4, NL5, NL6, NL7, NL8, NL9} and  $X_{\mu \nu \rho}=-X_{\nu \mu \rho}$ depends upon the spacetime torsion. Here, event $x'$ is connected  to event $x$ via a unique future directed timelike or null geodesic and we define the square of the proper length of this geodesic to be 2$\Omega$, where $\Omega$ is Synge's \emph{world function}~\cite{Sy}. In Eq.~\eqref{K5},  indices $\mu', \nu', \rho',...$ refer to event $x'$, while indices $\mu, \nu, \rho, ...$ refer to event $x$; moreover, 
\begin{equation}\label{K6}
\Omega_{\mu}(x, x'):=\frac{\partial \Omega}{\partial x^{\mu}}, \quad \Omega_{\mu'}(x, x'):=\frac{\partial \Omega}{\partial x'^{\mu'}}\,.
\end{equation}
For any bitensor, the covariant derivatives at $x$ and $x'$ commute~\cite{Sy}; that is,  $\Omega_{\mu \mu'}(x, x')=\Omega_{\mu' \mu}(x, x')$ is a dimensionless bitensor such that 
\begin{equation}\label{K7}
\lim_{x' \to x} \Omega_{\mu \mu'}(x, x')=-g_{\mu \mu'}(x)\,.
\end{equation}

To simplify matters further, we assume that $X_{\mu \nu \rho}$ is a \emph{linear} combination of the components of the torsion tensor, namely, 
\begin{equation}\label{K8}
X_{\mu \nu \rho}=\chi_{\mu \nu \rho}{}^{\alpha \beta \gamma}\,\mathfrak{C}_{\alpha \beta \gamma}\,.
\end{equation}
This relation is reminiscent of the local constitutive relation between $H_{\mu \nu}$ and $F_{\mu \nu}$ in electrodynamics~\cite{HO}. Various forms of Eq.~\eqref{K8} have been explored in  Ref.~\cite{NL9} and the relation that has been adopted for NLG is 
\begin{equation}\label{K9}
X_{\mu \nu \rho}= \mathfrak{C}_{\mu \nu \rho}+ p\,(\check{C}_\mu\, g_{\nu \rho}-\check{C}_\nu\, g_{\mu \rho})\,.
\end{equation}
Here, $p\ne 0$ is a constant dimensionless parameter and $\check{C}_\mu$ is the torsion pseudovector given by
\begin{equation}\label{K10}
\check{C}_\alpha=\frac{1}{3} E_{\alpha \beta \gamma \delta}\,\mathfrak{C}^{\beta \gamma \delta}\,,
\end{equation}
where $E_{\alpha \beta \gamma \delta}$ is the Levi-Civita tensor. 

It is interesting to express the main field equations of nonlocal gravity~\eqref{K2} as nonlocally modified Einstein's equations. To this end, Eq.~\eqref{J17} can be written as
\begin{eqnarray}\label{K11}
 {^0}G_{\mu \nu}=-\kappa E_{\mu \nu} +e_\mu{}^{\hat{\gamma}}\,g_{\nu \alpha}\,\frac{1}{\sqrt{-g}}~ \frac{\partial}{\partial x^\beta}\,\big(\sqrt{-g}\,\mathfrak{C}^{\alpha \beta}{}_{\hat{\gamma}}\big)\,.
\end{eqnarray}      
Let us now use Eq.~\eqref{K2} to write Eq.~\eqref{K11} as 
\begin{equation}\label{K12}
 {^0}G_{\mu \nu}+\Lambda\, g_{\mu \nu}+{\cal N}_{\mu \nu}-Q_{\mu \nu}=\kappa\, T_{\mu \nu}\,.
 \end{equation}
Here,  ${\cal N}_{\mu \nu}$ and $Q_{\mu \nu}$ are not in general symmetric tensors; moreover,  ${\cal N}_{\mu \nu}$  is given by
\begin{equation}\label{K13}
{\cal N}_{\mu \nu}:=g_{\nu \alpha}\, e_\mu{}^{\hat{\gamma}}\,\frac{1}{\sqrt{-g}}\,\frac{\partial}{\partial x^\beta}\,\Big(\sqrt{-g}\,N^{\alpha \beta}{}_{\hat{\gamma}}\Big)
\end{equation}
and $Q_{\mu \nu}:=\kappa\,(\mathcal{E}_{\mu \nu}-E_{\mu \nu})$ is a traceless tensor, namely, 
\begin{equation}\label{K14}
Q_{\mu \nu}:=C_{\mu \rho \sigma} N_{\nu}{}^{\rho \sigma}-\frac 14\, g_{\mu \nu}\,C_{ \delta \rho \sigma}N^{\delta \rho \sigma}\,.
\end{equation}
In Eq.~\eqref{K12}, we have sixteen field equations for the sixteen components of the gravitational potentials specified by our preferred tetrad field $e_\mu{}^{\hat{\alpha}}$. Nonlocal gravity is thus a tetrad theory that is invariant under the global Lorentz group and in which the Riemann curvature tensor and the Weitzenb\"ock torsion tensor both originate from the mass-energy content of the universe in accordance with Eq.~\eqref{K12}.

It remains to determine the constitutive kernel ${\cal K}(x, x')$. 
The structure of NLG implies that ${\cal K}$ could in general depend upon spacetime scalars at $x$ and $x'$ such as $\Omega_{\mu}(x, x')e^{\mu}{}_{\hat{\alpha}}(x)$ and $\Omega_{\mu'}(x, x')e^{\mu'}{}_{\hat{\alpha}}(x')$. We take the view that the functional form of the kernel should be determined from the observational data. A detailed discussion of this issue is contained in Refs.~\cite{NL1, NL2, NL3, NL4, NL5, NL6, NL7, NL8, NL9}.

The implications of this NLG theory have thus far been explored only in the linear regime. The investigation of the nonlinear regime of NLG remains a task for the future.

\section{Confrontation of NLG with Observation} 

In the general linear approximation of NLG, we can deal with problems regarding linearized gravitational radiation, bending of light and gravitational lensing as well as the Newtonian regime of nonlocal gravity. 

In the Newtonian limit of NLG, the Poisson equation for the gravitational potential $\Phi$ takes the form
\begin{equation}\label{L1}
\nabla^2\Phi = 4\pi G(\rho+\rho_D)\,, \qquad \rho_D(t, \mathbf{x})=\int q(\mathbf{x}-\mathbf{y}) \rho(t, \mathbf{y})d^3y\,,
\end{equation}
where $\rho_D$ has the interpretation of the density of ``dark matter" that is mimicked by nonlocality.  It turns out that in Eq.~\eqref{L1}, we can recover the phenomenological Tohline-Kuhn approach to modified gravity~\cite{T,K,B}. Indeed, for most situations of physical interest, the reciprocal kernel $q$ is a generalization of the Kuhn kernel~\cite{B} and is given by
\begin{equation}\label{L2}
 q=\frac{1}{4 \pi \lambda_0}\frac{(1+\mu r)}{r^2}e^{-\mu r}\,,
\end{equation}
where $r=|\mathbf{x}-\mathbf{y}|$ and $\lambda_0$ and $\mu$ are positive constant parameters such that $0<\mu \lambda_0<1$. It follows from Eqs.~\eqref{L1} and~\eqref{L2} that the attractive force of gravity acting on a point mass $m_1$  at $\mathbf{x}$  due to a point mass $m_2$ at $\mathbf{y}$ is given by
\begin{equation}\label{L3}
\mathbf{f}=-\frac{Gm_1m_2\,(\mathbf{x}-\mathbf{y})}{r^3} \Big[1+\alpha-\alpha (1+\frac{1}{2}\mu r)e^{-\mu r}\Big]\,,
\end{equation}
where $\alpha :=2/(\lambda_0 \mu)$ is a dimensionless parameter. In the framework of nonlocal gravity, Eq.~\eqref{L3} replaces the Newtonian inverse-square force law. A detailed analysis reveals that the gravitational physics of the solar system, spiral galaxies and clusters of galaxies can all be explained with Eq.~\eqref{L3} provided the parameters are chosen such that $\alpha \approx 11$, $\mu^{-1} \approx 17$ kpc and $\lambda_0 \approx 3$ kpc~\cite{NL8}. It is interesting to note that the nonlocally modified force law~\eqref{L3} consists of an enhanced attractive Newtonian part with $G \to G(1+\alpha)$ and a repulsive Yukawa part with a decay length of $\mu^{-1}$. These results are consistent with previous investigations~\cite{FPC, LL}.

Linearized gravitational radiation has been investigated within the framework of NLG~\cite{NL6, NL7, NL9}. The linearized gravitational waves with frequencies in the range that is currently of observational interest have wavelengths that are much shorter than 1 kpc; therefore, nonlocal effects in their generation and detection turn out to be negligible~\cite{NL6}. However, nonlocality does lead to the damping of gravitational waves as they propagate over cosmological distances, but the exponential damping time turns out to be longer than the age of the universe for gravitational waves 
of current observational interest~\cite{NL6, NL7}.

Nonlocal gravity has a galactic length scale of order 1 kpc; hence, NLG effects are generally negligible in systems with dimensions $\ll 1$ kpc, such as planetary systems or binary pulsars. On the other hand, it would be most interesting to detect the influence of nonlocality in such systems. In this connection, the NLG-induced periastron precession must be mentioned. The effect is at present buried in the noise; e.g., it is retrograde and the ratio of its magnitude for Mercury to Einstein's precession is about $10^{-3}$~\cite{NL3, NL8, LI, LP}. However, future observations of binary pulsars may lead to the
 detection of this effect~\cite{LLY}.

It remains to confront the predictions of NLG with gravitational lensing observations~\cite{NL9}. The implications of NLG for structure formation in cosmology also remain a task for the future.

\section{Discussion}

Among the fundamental interactions, gravitation has the unique feature of universality. It may also be history dependent. In this paper, the main aspects of the recent nonlocal generalization of Einstein's theory of gravitation have been briefly described. If it turns out that nonlocal gravity is supported by observational data, this may lead to a deeper understanding of the gravitational interaction and provide a clue towards its eventual quantization~\cite{BeRe}. 

\begin{acknowledgments}
I am grateful to Jos\'e~W.~Maluf for valuable discussions. 
\end{acknowledgments}

\appendix{}

\section{Measurement of Weitzenb\"ock's Torsion}\label{appA}

Torsion and curvature are the two fundamental differential geometric notions associated with a linear connection, or the corresponding covariant differentiation, on a manifold. Torsion has to do with the lack of symmetry of the connection and curvature is related to the lack of commutativity of covariant differentiation~\cite{BEE}. The Gaussian curvature of a surface is its most significant property and is easy to visualize~\cite{Ha}. In general, spacetime curvature can be \emph{operationally} defined, for instance, via geodesic deviation (the Jacobi equation and its generalizations) or via parallel vector fields (holonomy). Torsion, on the other hand, can be visualized as the failure of a parallelogram to close, a concept related to the presence of dislocations in continuous media~\cite{HO, MUF}. However, in contrast to the case of curvature, there is no general operational definition for the torsion of spacetime. In particular, the torsion tensor apparently has no relation with the torsion of a curve in space~\cite{Hi}.

It appears that the measurement of spacetime torsion depends upon the physical theory in which torsion plays a significant role. In the Poincar\'e gauge theory of gravitation, for instance, Cartan's torsion couples to intrinsic spin and its measurement has been discussed in that context---see~\cite{He, La, HOP} and the references cited therein. However, the teleparallel equivalent of general relativity, GR$_{||}$, is considered to be a degenerate limit of the Poincar\'e gauge theory~\cite{BH} and requires a separate treatment that is the main subject of this section. Indeed, spin effects in GR$_{||}$ are expected to be essentially the same as in GR~\cite{AP, MaOb}.

Within the framework of teleparallelism, the metric is connected to the preferred parallel frame field via orthonormality; furthermore, the Weitzenb\"ock torsion is naturally related to the Riemannian curvature of spacetime. The elements necessary for the establishment of metric geometry, namely, infinitesimal rods, clocks, light signals, etc., may then be employed to provide an \emph{indirect} operational definition of  Weitzenb\"ock's torsion. This is illustrated below via a specific example involving a frame field in an arbitrary Fermi coordinate system. 

Consider a gravitational field in the extended GR framework; that is, we allow for nonlocality,  a possibility that is immaterial for our considerations. It is possible to establish an extended quasi-inertial Fermi normal coordinate system $X^\mu$ in a world tube along the path of a  test observer following a geodesic in this curved spacetime. These coordinates are scalar invariants by construction and are indispensable for the interpretation of measurements in GR. More specifically, imagine a congruence of future directed timelike geodesics representing the motion of free test observers  in an arbitrary 
gravitational field.  In this
congruence, we choose a reference observer ${\cal O}$  that follows a world line ${\bar x}^\mu (\tau)$ and carries, via the Levi-Civita connection, an orthonormal parallel-propagated tetrad frame $\lambda^{\mu}{}_{\hat{\alpha}}(\tau )$ along its path. Here, $\tau$ is the reference observer's proper time; moreover,  $\lambda^\mu{}_{\hat{0}}=d{\bar x}^\mu/d\tau$ is a unit timelike vector tangent to the world line of the
observer ${\cal O}$ and is its local temporal axis, while $\lambda^{\mu}{}_{\hat{i}}$, $i=1,2,3$, form its local spatial frame. We wish to introduce a geodesic coordinate system in the neighborhood of the world line of ${\cal O}$---see~\cite{Sy, CM1, CM2, CM3} and the references cited therein. At each event $Q(\tau)$ along ${\bar x}^\mu (\tau)$, the class of spacelike geodesics orthogonal to ${\bar x}^\mu (\tau)$ 
form a local hypersurface. Let $P$ be an event with coordinates $x^\mu$ on this hypersurface such that there is a \emph{unique} spacelike geodesic that connects $Q$ to $P$. The Fermi coordinates of $P$ are $X^\mu=(cT, X^i)$, which are defined by
\begin{equation}\label{A1}
T=\tau\,, \qquad X^i=\sigma\, \xi^\mu\, \lambda_{\mu}{}^{\hat{i}}(\tau)\,,
 \end{equation}
where $\xi^\mu$ is the unit vector at $Q(\tau)$ that is tangent to the unique spacelike geodesic segment from $Q$ to $P$ and $\sigma$ is the proper length of this segment. The reference observer ${\cal O}$ thus permanently resides at the spatial origin of the Fermi coordinate system.

In general, the coordinate transformation $x^\mu \mapsto X^\mu$ can only be specified implicitly; therefore, it becomes necessary  to have Taylor expansions in powers of the spatial distance away from the reference world line. Indeed, the spacetime metric in Fermi coordinates is given by
\begin{align}\label{A2} 
g_{00}&=-1-\widehat{R}_{0i0j}(T)X^iX^j+\ldots ,\\
\label{A3} g_{0i}&=-\frac{2}{3}\widehat{R}_{0jik}(T)X^jX^k+\ldots ,\\
\label{A4} g_{ij}&=\delta_{ij}-\frac{1}{3}\widehat{R}_{ikjl}(T)X^kX^l+\ldots ,
\end{align}
where $\widehat{R}_{\alpha \beta \gamma \delta}(T)$ is the projection of the 
Riemann curvature tensor evaluated along the reference geodesic, where $T=\tau$ and ${\bf X}=0$, on the orthonormal tetrad frame of the
reference observer ${\cal O}$; that is,
\begin{equation}\label{A5} 
\widehat{R}_{\alpha \beta \gamma \delta}(T):=\,^{0}R_{\mu\nu \rho
\sigma}\,\lambda^\mu{}_{\hat{\alpha}}\,\lambda^\nu{}_{\hat{\beta}}\,\lambda^\rho{}_{\hat{\gamma}}\,\lambda^\sigma{}_{\hat{\delta}}\,.
\end{equation}
The Fermi coordinates are admissible in a finite cylindrical region about the world line of ${\cal O}$; moreover, the radius of the cylinder is essentially characterized by the radius of curvature of spacetime.

In the quasi-inertial Fermi coordinate system, the spacetime metric  may be written as $g_{\mu \nu}=\eta_{\mu \nu}+h_{\mu \nu}(X)$, which is the metric of a perturbed Minkowski spacetime. It proves useful to characterize this perturbation within the context of the spacetime curvature approach to gravitoelectromagnetism (GEM), namely, $h_{00} :=-2\Phi$, $h_{0i} := -2 {\cal A}_i$ and $h_{ij} :=-2\Sigma_{ij}$, where $\Phi$ is the generalization of the Newtonian potential in this case, so that $-\Phi$ is the gravitoelectric potential, and $\boldsymbol{{\cal A}}$ is the gravitomagnetic vector potential~\cite{M9}. More explicitly, we set
\begin{equation}\label{A6} 
\Phi=\frac{1}{2}\widehat{R}_{0i0j}(T)X^iX^j\,, \quad {\cal A}_i=\frac{1}{3}\widehat{R}_{0jik}(T)X^jX^k\,, \quad \Sigma_{ij}=\frac{1}{6}\widehat{R}_{ikjl}(T)X^kX^l\,,
\end{equation}
where only the dominant terms in the perturbation have been retained. The gravitoelectric field, $\boldsymbol{{\cal E}}=\boldsymbol{\nabla} \Phi$, and the gravitomagnetic field, $\boldsymbol{{\cal B}}=\boldsymbol{\nabla} \times \boldsymbol{{\cal A}}$\,, are then given by
\begin{align}\label{A7} 
{\cal E}_i(T ,{\bf X})&=\widehat{R}_{0i0j}(T) X^j+\ldots ,\\
\label{A8} {\cal B}_i(T ,{\bf X})&=-\frac{1}{2}\epsilon_{ijk}\widehat{R}^{jk}{}_{0l}(T )X^l +\ldots .
\end{align}
Henceforth, we will ignore all higher-order terms in the perturbation and note that with this simplification the gravitoelectric field becomes directly proportional to the ``electric" components of the of the Riemann curvature tensor $\widehat{R}_{0i0j}$ and similarly the gravitomagnetic 
field is directly proportional to the ``magnetic''
components of the Riemann curvature tensor $\widehat{R}_{0ijk}$~\cite{Mat}. Moreover, the spatial part of the metric perturbation, $\Sigma_{ij}=\Sigma_{ji}$, is likewise proportional to the spatial components of the curvature $\widehat{R}_{ijkl}$.

Let us now consider the class of observers that are all at \emph{rest} in this gravitational field and carry orthonormal tetrads that have essentially the same orientation as the Fermi coordinate system. This class includes of course our reference observer ${\cal O}$. The orthonormal tetrad frame of these preferred observers can be expressed in $(cT, X^i)$ coordinates as 
\begin{align}
\label{A9} e^{\mu}{}_{\hat{0}}&= (1-\Phi,\, 0,\, 0,\, 0)\,,\\
\label{A10} e^{\mu}{}_{\hat{1}}&=(-2{\cal A}_1,\, 1+\Sigma_{11},\, 0,\, 0)\,,\\
\label{A11} e^{\mu}{}_{\hat{2}}&=(-2{\cal A}_2,\, 2\,\Sigma_{12},\, 1+\Sigma_{22},\, 0)\,,\\
\label{A12} e^{\mu}{}_{\hat{3}}&=(-2{\cal A}_3,\, 2\,\Sigma_{13},\, 2\,\Sigma_{23},\, 1+\Sigma_{33})\,.
\end{align}
As expected, $e^{\mu}{}_{\hat{\alpha}}$ reduces to $\delta^\mu_\alpha$ either in the absence of spacetime curvature or along the reference geodesic, where $\mathbf{X}=0$. To have a globally parallel frame field for the preferred observers, we introduce the Weitzenb\"ock connection and the associated torsion tensor as in Eqs.~\eqref{J8} and~\eqref{J10}, respectively. The components of the torsion tensor as measured by the preferred observers themselves can be obtained from projecting the torsion tensor on the tetrad frames of these observers, namely, 
\begin{equation}\label{A13}
 C_{\hat{\alpha} \hat{\beta}}{}^{\hat{\gamma}}=e^{\mu}{}_{\hat{\alpha}}\,e^{\nu}{}_{\hat{\beta}}\Big(\partial_{\mu}e_{\nu}{}^{\hat{\gamma}}-\partial_{\nu}e_{\mu}{}^{\hat{\gamma}}\Big)\,.
\end{equation}

In general, other classes of observers may be considered that are all related to the static observers by constant boosts and rotations of their tetrads~\eqref{A9}--\eqref{A12}. The corresponding measured torsion tensor is then related to Eq.~\eqref{A13} by global Lorentz transformations. 

Equation~\eqref{A13} can be much simplified if we recall that $e^{\mu}{}_{\hat{\alpha}}$ differs from $\delta^\mu_\alpha$ by a small perturbation, which we denote by $-\psi^{\mu}{}_{\alpha}$, since the distinction between holonomic and anholonomic indices disappears when dealing with this perturbation tensor. In fact, we find that  
\begin{equation}\label{A14}
(\psi_{\alpha \beta})=-\, \left[
\begin{array}{cccc}
 \Phi & 2{\cal A}_1 & 2{\cal A}_2 & 2{\cal A}_3\cr
 0 & \Sigma_{11} &  2\Sigma_{12} &  2\Sigma_{13}\cr
 0 & 0 &  \Sigma_{22} &  2\Sigma_{23}\cr
 0 & 0 & 0&  \Sigma_{33}\cr 
 \end{array}
 \right]\,.
\end{equation}
Therefore, the measured components of the torsion tensor in this linear approximation scheme may be expressed as 
\begin{equation}\label{A15}
 C_{\alpha \beta}{}^\gamma=\partial_{\alpha}\psi^{\gamma}{}_{\beta}-\partial_{\beta}\psi^{\gamma}{}_{\alpha}\,.
\end{equation}
In this relation, for each $\gamma=0, 1, 2, 3$, we have an antisymmetric tensor that has ``electric" and ``magnetic" components in analogy with the electromagnetic field tensor. Indeed, for $\gamma=0$, the electric part corresponds to the gravitoelectric field~\eqref{A7} and the magnetic part is twice the gravitomagnetic field~\eqref{A8} in our linear approximation scheme; that is, 
\begin{equation}\label{A16}
 C_{0i}{}^0=-{\cal E}_i\,,  \qquad  C_{ij}{}^0=2 \epsilon_{ijk}{\cal B}^k\,.
 \end{equation}
Moreover, for $\gamma=1, 2, 3$, the electric parts only involve terms of higher order and can be ignored, so that
\begin{equation}\label{A17}
 C_{0i}{}^j=0\,.
 \end{equation}
However, the corresponding magnetic parts depend upon the spatial components of the curvature and we find that for $\gamma=1$, 
\begin{equation}\label{A18}
 C_{23}{}^1=\widehat{R}_{231i}X^i\,, \quad  C_{31}{}^1=\frac{2}{3}\widehat{R}_{311i}X^i\,, \quad C_{12}{}^1=\frac{2}{3}\widehat{R}_{121i}X^i\,.
 \end{equation}
Similarly, for $\gamma=2$,
\begin{equation}\label{A19}
 C_{23}{}^2=\frac{2}{3}\widehat{R}_{232i}X^i\,, \quad  C_{31}{}^2=\frac{1}{3}(\widehat{R}_{312i}-\widehat{R}_{123i})X^i\,, \quad C_{12}{}^2=\frac{1}{3}\widehat{R}_{122i}X^i\,,
 \end{equation}
and for $\gamma=3$, 
\begin{equation}\label{A20}
 C_{23}{}^3=\frac{1}{3}\widehat{R}_{233i}X^i\,, \quad  C_{31}{}^3=\frac{1}{3}\widehat{R}_{313i}X^i\,, \quad C_{12}{}^3=0\,.
 \end{equation}
Using the antisymmetry of the torsion tensor  in its first two indices, all of the components of $C_{\alpha \beta}{}^{\gamma}$ can be obtained from Eqs.~\eqref{A16}--\eqref{A20}; in fact, \emph{all} of the components of the curvature tensor are involved in the determination of the torsion tensor.  Moreover, it is straightforward to use these components to compute the elements of the torsion tensor that are irreducible under the global Lorentz group, namely, the torsion vector, the torsion pseudovector and the reduced torsion tensor.

 We note that the torsion tensor completely vanishes  along the reference \emph{geodesic} $(\mathbf {X}=0)$. This means that the contorsion tensor and hence the Weitzenb\"ock connection vanish as well along the reference geodesic. We thus have a generalization of Fermi's result, namely, that the components of both the Levi-Civita and Weitzenb\"ock connections can be made to vanish along a timeline geodesic by a proper choice of coordinates and preferred tetrad frames. The investigation of the case of an \emph{accelerated} reference path is beyond the scope of this work.  The corresponding generalization of the Fermi normal coordinate system in GR  has been treated in Refs.~\cite{M10, NZ}. The limit of vanishing curvature is considered in Appendix B.
 
The preferred observers that are all at rest in the Fermi system are generally accelerated with an acceleration vector that is given in Fermi coordinates by $A_\mu = (0, \mathcal{E}_i)$, which is valid to linear order in $\mathbf{X}$. The geodesic equation of motion of a free test particle in the Fermi system with velocity $\mathbf{V}=d\,\mathbf{X}/d\,T$ can be written as
\begin{equation}\label{A21}
 \frac{d^2X^i}{d\,T^2}+ \widehat{R}_{0i0j}X^j + 2 \widehat{R}_{ikj0}V^kX^j+\Big( 2 \widehat{R}_{0kj0}V^iV^k+ \frac{2}{3}\widehat{R}_{ikjl}V^kV^l+  \frac{2}{3}\widehat{R}_{0kjl}V^iV^kV^l \Big)\,X^j=0\,.
 \end{equation}
This is the \emph{generalized Jacobi equation}~\cite{CM1}, where tidal accelerations are given to first order in $\mathbf{X}$. It thus appears that the measured components of the torsion tensor~\eqref{A16}--\eqref{A20} are closely related to relativistic tidal accelerations. 
 
The curvature components $\widehat{R}_{\alpha \beta \gamma \delta}$ and the spatial Fermi coordinates $\bf X$ can be measured as in standard GR; therefore, the Weitzenb\"ock torsion, like tidal acceleration, has dimensions of (length)$^{-1}$ and can be measured indirectly via Eqs.~\eqref{A16}--\eqref{A20}. By introducing a quasi-inertial Fermi coordinate system and a preferred class of observers in an arbitrary curved spacetime region, we have demonstrated that the gravitational field as given by the torsion tensor has physical significance, since in principle it can be measured.  The connection between this approach to torsion and the lack of closure of parallelograms deserves further investigation~\cite{MUF}.

\section{Weitzenb\"ock's torsion in Minkowski spacetime}\label{appB}

In teleparallelism as in GR, gravitation is characterized by the Riemannian curvature of spacetime; thus, so long as $^{0}R_{\mu \nu \rho \sigma}$ is nonzero, the Weitzenb\"ock  torsion tensor, which represents  the \emph{gravitational field} in teleparallelism, is nonzero as well. As demonstrated in Appendix A, the torsion tensor is physically measurable and its components are closely associated with the Riemannian curvature of spacetime.  On the other hand, if $^{0}R_{\mu \nu \rho \sigma}=0$, then in flat spacetime  Weitzenb\"ock's torsion tensor loses its gravitational significance. In arbitrary systems of admissible coordinates in Minkowski spacetime, the torsion tensor vanishes only for inertial observers that are at rest in a global inertial frame and have orthonormal tetrad axes that are all parallel to the standard Cartesian coordinate axes of the global inertial frame; otherwise, the torsion tensor is nonzero. Thus accelerated observers are endowed with torsion; similarly, torsion is nonzero for inertial observers that are static in a global inertial frame but have spatial frames that vary in space~\cite{HS, MVRN}. To illustrate the latter possibility, consider, for instance, a global inertial frame and static \emph{inertial} observers with orthonormal tetrads such that their spatial frames are all along the spherical polar coordinate axes; in this case, the torsion tensor has spatial components that do not vanish~\cite{MVRN}. The purpose of this Appendix is to show how  Weitzenb\"ock's torsion is related to the \emph{acceleration tensor} of  observer families in Minkowski spacetime.

Imagine an accelerated observer in a \emph{global inertial frame} in Minkowski spacetime. The observer follows the reference world line ${\bar x}^\mu (\tau)$, where $\tau$ is its proper time; moreover, it carries along this path an orthonormal tetrad frame $\lambda^{\mu}{}_{\hat{\alpha}}(\tau )$, where  $\lambda^\mu{}_{\hat{0}}=d{\bar x}^\mu/d\tau$ is its unit  temporal axis and $\lambda^{\mu}{}_{\hat{i}}$\,,\, $i=1,2,3$, constitute its local spatial frame. Following the idea of a moving frame field, we have
\begin{equation}\label{B1} 
\frac{d\lambda^\mu{}_{\hat{\alpha}}}{d\tau} =\phi_{\hat{\alpha}}{}^{\hat{\beta}}(\tau)~ 
\lambda^\mu{}_{\hat{\beta}}\,.
\end{equation}
Here, $\phi_{\hat{\alpha} \hat{\beta}}$ is the antisymmetric acceleration tensor such that $\phi_{\hat{0} \hat{i}}=a_i$ and $\phi_{\hat{i} \hat{j}} = \epsilon_{ijk}\,\omega^k$ are respectively the tetrad components of the reference observer's translational acceleration of its world line and the rotational angular velocity of its spatial frame with respect to the local nonrotating (i.e., Fermi-Walker transported)  frame. Let us now consider a geodesic system of coordinates $X^\mu=(cT, X^i)$ established along the world line of the reference observer. Given any
event $\tau$ along ${\bar x}^\mu (\tau)$, the straight
spacelike geodesic lines orthogonal to the reference observer's world line span a 
hyperplane that is in fact the three-dimensional Euclidean space. For an event on this hyperplane with inertial coordinates $x^\mu$, the relationship between $x^\mu$ and $X^\alpha$ is given by
\begin{equation}\label{B2} 
\tau =X^0\,, \qquad  x^\mu =\bar{x}^\mu(\tau)+X^i\lambda^\mu{}_{\hat{i}}(\tau)\,.
\end{equation}
 It follows from Eqs.~\eqref{B1} and~\eqref{B2} that 
\begin{equation}\label{B3} 
dx^\mu =({\cal P} \lambda^\mu{}_{\hat{0}} + {\cal Q}^j \lambda^\mu{}_{\hat{j}})\,dX^0+\lambda^\mu{}_{\hat{i}}\,dX^i\,,
\end{equation} 
where 
\begin{equation}\label{B4} 
{\cal P}=1+\mathbf{a} \cdot \mathbf{X}\,, \qquad  {\cal Q}_{i}= (\boldsymbol{\omega} \times \mathbf{X})_i\,.
\end{equation}  
It is then simple to show that the Minkowski metric $\eta _{\mu\nu}\,dx^\mu \otimes dx^\nu$ with respect to 
the new geodesic coordinate system can be written as $g_{\mu\nu}\,dX^\mu \otimes dX^\nu$, where
\begin{equation}\label{B5}
 g_{00}=-{\cal P}^2+{\cal Q}^2\,, \quad  g_{0i}={\cal Q}_i\,, \quad g_{ij}=\delta_{ij}\,.
\end{equation}
The new coordinates are admissible in a cylindrical spacetime region around $\bar{x}^\mu(\tau)$ so long as $g_{00}<0$. For  a detailed discussion of this condition as well as the properties of the boundary hypersurface, see Ref.~\cite{M11}; moreover, a general discussion of the inertial effects  in geodesic coordinates and further references are contained in Ref.~\cite{CM2}. 

Consider now the class of accelerated observers that are at \emph{rest} in space in the geodesic coordinate system $X^\mu=(cT, X^i)$, and carry spatial frames that are all nearly aligned with the spatial frame of the reference observer. To simplify matters, we will assume that $|\mathbf{a} \cdot \mathbf{X}| \ll 1$ and $|{\cal Q}_{i}| \ll 1$; then, to first order in these small quantities---i.e., to linear order in $\mathbf{X}$---the orthonormal tetrad frames of these preferred observers are given in $X^\mu=(cT, X^i)$ coordinates by
\begin{align}
\label{B6} e^{\mu}{}_{\hat{0}}&= (1-\mathbf{a} \cdot \mathbf{X},\, 0,\, 0,\, 0)\,,\\
\label{B7} e^{\mu}{}_{\hat{1}}&=({\cal Q}_1,\, 1,\, 0,\, 0)\,,\\
\label{B8} e^{\mu}{}_{\hat{2}}&=({\cal Q}_2,\, 0,\, 1,\, 0)\,,\\
\label{B9} e^{\mu}{}_{\hat{3}}&=({\cal Q}_3,\, 0,\, 0,\, 1)\,.
\end{align}
As expected, for $\mathbf{X}=0$, this tetrad system coincides with that of the reference observer in geodesic coordinates; i.e., the reference observer is naturally a preferred observer as well. 

Our linear approximation scheme is such that $e^{\mu}{}_{\hat{\alpha}}$ differs from $\delta^\mu_\alpha$ by quantities whose magnitudes are very small compared to unity. Thus let $e^{\mu}{}_{\hat{\alpha}}=\delta^\mu_\alpha - \psi^\mu{}_\alpha$ as before;  then, 
\begin{equation}\label{B10} 
\psi_{00}=-\mathbf{a} \cdot \mathbf{X}\,,\quad \psi_{0i}={\cal Q}_{i}\, \quad \psi_{i \alpha}=0\,.
\end{equation}  
In the linear approximation, the components of the torsion tensor are then given, as before, by Eq.~\eqref{A15}. It follows from Eq.~\eqref{B10} that, to lowest order, the torsion tensor is the same for the class of preferred observers and is thus only a function of the temporal variable $T$; that is, 
\begin{equation}\label{B11} 
C_{0i}{}^{0}=-a_i(T)\,, \qquad C_{ij}{}^{0}=-2\epsilon_{ijk}\,\omega^k(T)\,,
\end{equation}  
while $C_{\alpha \beta}{}^{i}=0$. Thus the torsion tensor contains essentially the same information as the acceleration tensor. It is interesting to compare and contrast this inertial result involving acceleration with the gravitational case involving \emph{tidal} acceleration in Appendix A.


\begin{thebibliography}{99}

\bibitem{Ei} 
A.~Einstein, \textit{The Meaning of Relativity}  (Princeton University Press, Princeton, NJ, 1955).

\bibitem{Mas} 
B.~Mashhoon, Phys. Rev. Lett. \textbf{61}, 2639 (1988);

B.~Mashhoon, Phys. Lett. A \textbf{143}, 176 (1990);

B.~Mashhoon, ``The Hypothesis of Locality in Relativistic Physics", Phys. Lett. A \textbf{145}, 147 (1990).

\bibitem{BM}
B.~Mashhoon, ``Nonlocal Theory of Accelerated Observers", Phys. Rev. A {\bf 47}, 4498 (1993).

\bibitem{M1} 
B.~Mashhoon, ``Nonlocal Special Relativity", 
Ann. Phys. (Berlin) {\bf 17}, 705 (2008) [arXiv: 0805.2926 [gr-qc]].

\bibitem{NL1} F.~W.~Hehl and B.~Mashhoon, ``Nonlocal Gravity Simulates Dark Matter", 
Phys. Lett. B {\bf 673}, 279 (2009) [arXiv: 0812.1059 [gr-qc]].

\bibitem{NL2} F.~W.~Hehl and B.~Mashhoon, ``Formal Framework for a Nonlocal Generalization of Einstein's Theory of Gravitation", 
Phys. Rev. D {\bf 79}, 064028 (2009) [arXiv: 0902.0560 [gr-qc]].

\bibitem{NL3} H.-J.~Blome, C.~Chicone, F.~W.~Hehl and B.~Mashhoon, ``Nonlocal Modification of Newtonian Gravity", Phys. Rev. D {\bf 81}, 065020 (2010) [arXiv:1002.1425 [gr-qc]].
    
\bibitem{NL4} B.~Mashhoon, ``Nonlocal Gravity", in \emph{Cosmology and Gravitation}, edited by M. Novello and S. E. Perez Begliaffa (Cambridge Scientific Publishers, Cambridge, UK, 2011), pp.~1--9 [arXiv:1101.3752 [gr-qc]].

\bibitem{NL5} C.~Chicone and B.~Mashhoon, ``Nonlocal Gravity: Modified Poisson's Equation", J. Math. Phys. {\bf  53}, 042501 (2012) [arXiv:1111.4702 [gr-qc]].

\bibitem{NL6} C.~Chicone and B.~Mashhoon, ``Linearized Gravitational Waves in Nonlocal General Relativity", Phys. Rev. D {\bf 87}, 064015 (2013) [arXiv:1210.3860 [gr-qc]].

\bibitem{NL7}
B.~Mashhoon, ``Nonlocal Gravity: Damping of Linearized Gravitational Waves", Classical Quantum Gravity {\bf 30}, 155008 (2013) [arXiv:1304.1769 [gr-qc]].

\bibitem{NL8} S.~Rahvar and B.~Mashhoon, ``Observational Tests of Nonlocal Gravity: Galaxy Rotation Curves and Clusters of Galaxies", Phys. Rev. D {\bf 89}, 104011 (2014) [arXiv:1401.4819 [gr-qc]].
    
\bibitem{NL9} B.~Mashhoon, ``Nonlocal Gravity: The General Linear Approximation", Phys. Rev. D {\bf 90}, 124031(2014) [arXiv:1409.4472 [gr-qc]].    

\bibitem{BR} 
N.~Bohr and L.~Rosenfeld, K. Dan. Vidensk. Selsk. Mat. Fys. Medd. \textbf{12}, 8 (1933) [ translated in \textit{Quantum Theory and Measurement}, edited by J.~A.~Wheeler and W.~H.~Zurek (Princeton University Press, Princeton, NJ, 1983)];
	 
N.~Bohr and L.~Rosenfeld, Phys. Rev. \textbf{78}, 794 (1950).

\bibitem{BH}
M.~Blagojevi\'c and F.~W.~Hehl, editors, {\it Gauge Theories of Gravitation}
(Imperial College Press, London, 2013).


\bibitem{AP}
R.~Aldrovandi and J.~G.~Pereira, 
\textit{Teleparallel Gravity: An Introduction} (Springer, New York, 2013).

\bibitem{Ma}
J.~W.~Maluf, ``The Teleparallel Equivalent of General Relativity", Ann. Phys. (Berlin) {\bf 525}, 339 (2013).


\bibitem{TW}
N.~C.~Tsamis and R.~P.~Woodard, ``A Caveat on Building Nonlocal Models of Cosmology", J. Cosmol. Astropart. Phys. 09 (2014) 008 [arXiv:1405.4470 [astro-ph.CO]].

\bibitem{BMMS}
F.~ Briscese, A.~Marcian\`o, L.~Modesto and E.~N.~Saridakis, ``Inflation in (Super-)Renormalizable Gravity", Phys. Rev. D \textbf{87}, 083507 (2013) [arXiv:1212.3611 [hep-th]].


\bibitem{Sy}
 J.~L.~Synge, {\it Relativity: The General Theory} (North-Holland, Amsterdam, 1971).
 
 \bibitem{BCM}
D.~Bini, C.~Chicone and B.~Mashhoon, ``Spacetime Splitting, Admissible Coordinates, and Causality", Phys. Rev. D {\bf 85}, 104020 (2012).

 
 \bibitem{We}
R.~Weitzenb\"ock, {\it Invariantentheorie} (Noordhoff, Groningen, 1923).

\bibitem{M8}
B.~Mashhoon,
{\it Phys. Lett. A} {\bf 122}, 299 (1986). 

\bibitem{LB}
Ll.~Bel, ``Connecting Connections", arXiv:0805.0846 [gr-qc].


\bibitem{Ein}
A.~Einstein, Math. Ann. {\bf 102}, 685 (1930).

\bibitem{Mo}
C.~M{\o}ller, K. Dan. Vidensk. Selsk. Mat. Fys. Skr. \textbf{1}, 10 (1961).

\bibitem{PP}
C.~Pellegrini and J.~Pleba\'nski, K. Dan. Vidensk. Selsk. Mat. Fys. Skr. \textbf{2}, 4 (1963).

 \bibitem{Ja}
 J.~D.~Jackson, {\it Classical Electrodynamics} (Wiley, Somerset, NJ, 1999), 3rd ed.
 
\bibitem{LaLi}
L.~D.~Landau and E.~M.~Lifshitz, {\it Electrodynamics of Continuous Media} (Pergamon, New York, 1960).

\bibitem{HO}  
F.~W.~Hehl and Yu.~N.~Obukhov, {\it Foundations of
    Classical Electrodynamics: Charge, Flux, and Metric} (Birkh\"auser, Boston, MA, 2003).

\bibitem{T}
 J.~E.~Tohline, in {\it IAU Symposium 100,
    Internal Kinematics and Dynamics of Galaxies,} edited by
  E.~Athanassoula (Reidel, Dordrecht, 1983), p.~205;
 J.~E.~Tohline, Ann. N.Y. Acad. Sci. {\bf 422}, 390 (1984). 
 
\bibitem{K} 
J.~R.~Kuhn and L.~Kruglyak,   Astrophys.\ J. {\bf 313}, 1 (1987).

\bibitem{B}
 J.~D.~Bekenstein, in {\it Second Canadian
    Conference on General Relativity and Relativistic Astrophysics}, edited by
  A.~Coley, C.~Dyer and T.~Tupper (World Scientific, Singapore,
  1988), p.~68.


 \bibitem{RF} 
V.~C.~Rubin and W.~K.~Ford, Astrophys. J.  {\bf 159}, 379 (1970).  

\bibitem{RW}
 M.~S.~Roberts and R.~N.~Whitehurst, Astrophys.
  J.  {\bf 201}, 327 (1975).

\bibitem{SR} 
Y.~Sofue and V.~Rubin, 
Annu. Rev. Astron. Astrophys. {\bf 39}, 137 
    (2001).
     
\bibitem{FPC}
J.~C.~Fabris and J.~Pereira Campos, Gen. Relativ. Gravit. {\bf 41}, 93 (2009).

\bibitem{LL}
S.~Little and M.~Little, Classical Quantum Gravity {\bf 31}, 195008 (2014).

\bibitem{LI}
L.~Iorio, J. High Energy Phys. 05 (2012) 073.

\bibitem{LP}
D.~M.~Lucchesi and R.~Peron, Phys. Rev. D {\bf 89}, 082002 (2014).

\bibitem{LLY}
C.~Lu \emph{et al.}, RAA (Research in Astronomy and Astrophysics) {\bf 14}, 1301 (2014).

\bibitem{BeRe}
D.~Becker and M.~Reuter, arXiv:1407.5848  [hep-th].

\bibitem{BEE}
 J.~K.~Beem, P.~E.~Ehrlich and K.~L.~Easley, {\it Global Lorentzian Geometry} (Marcel Dekker, New York, 1996), 2nd ed.


\bibitem{Ha} 
E.~Harrison, {\it Cosmology} (Cambridge University Press, Cambridge, UK, 2000), 2nd ed. 


\bibitem{MUF}
J.~W.~Maluf, S.~C.~Ulhoa and F.~F.~Faria, ``Pound-Rebka Experiment and Torsion in the Schwarzschild Spacetime", Phys. Rev. D {\bf 80}, 044036 (2009).

\bibitem{Hi}
N.~J.~Hicks, {\it Notes on Differential Geometry} (D. Van Nostrand, Princeton, NJ, 1965).

\bibitem{He}  
F.~W.~Hehl, Phys. Lett. A {\bf  36}, 225 (1971).

\bibitem{La}  
C.~L\"ammerzahl, Phys. Lett. A {\bf  228}, 223 (1997).

\bibitem{HOP}  
F.~W.~Hehl, Yu.~N.~Obukhov and D.~Puetzfeld, Phys. Lett. A {\bf  377}, 1775 (2013)  [arXiv:1304.2769 [gr-qc]].

\bibitem{MaOb}
B.~Mashhoon, and Yu.~N.~Obukhov, Phys. Rev. D {\bf 88}, 064037 (2013)  [arXiv:1307.5470 [gr-qc]].


\bibitem{CM1} 
C.~Chicone and B.~Mashhoon,  Classical Quantum Gravity
{\bf 19}, 4231 (2002)  [arXiv:gr-qc/0203073].

\bibitem{CM2} 
C.~Chicone and B.~Mashhoon,  Classical Quantum Gravity 
{\bf 22}, 195 (2005)  [arXiv:gr-qc/0409017].

\bibitem{CM3}
 C.~Chicone and B.~Mashhoon, Phys. Rev. D 
{\bf 74}, 064019 (2006)  [arXiv:gr-qc/0511129].

\bibitem{M9}
B.~Mashhoon, ``Gravitoelectromagnetism: A Brief Review", 
in \emph{The Measurement of Gravitomagnetism:
A Challenging Enterprise}, edited by L. Iorio (Nova Science,
New York, 2007), Chap. 3, pp. 29--39  [arXiv:gr-qc/0311030].

\bibitem{Mat} 
A.~Matte,  Canadian J. Math. {\bf 5}, 1 (1953).

\bibitem{M10}
B.~Mashhoon, Astrophys.\ J. {\bf 216}, 591 (1977).

\bibitem{NZ}
W.-T.~Ni and M.~Zimmermann, Phys. Rev. D {\bf 17}, 1473 (1978).

\bibitem{HS}
K.~Hayashi and T.~Shirafuji, ``New General Relativity", Phys. Rev. D {\bf 19}, 3524 (1979).

\bibitem{MVRN}
J.~W.~Maluf, M.~V.~O.~Veiga and J.~F.~da Rocha-Neto, ``Regularized Expression for the Gravitational Energy-Momentum in Teleparallel Gravity and the Principle of Equivalence", Gen. Relativ. Gravit. {\bf 39}, 227 (2007).


\bibitem{M11} 
B.~Mashhoon, in {\it Advances in General Relativity and Cosmology}, 
edited by G.~ Ferrarese (Pitagora, Bologna,
2003), pp. 323-334 [arXiv:gr-qc/0301065].


\end{thebibliography}
\end{document}